\documentstyle[12pt]{article}
\makeatletter
 
 \@addtoreset{equation}{section}
\makeatother

\newcommand{\sect}[1]{\section{#1}\setcounter{equation}{0}}

\newcommand{\beq}{\begin{equation}}
\newcommand{\eeq}{\end{equation}}
\newcommand{\beqs}{\begin{eqnarray}}
\newcommand{\eeqs}{\end{eqnarray}}

\begin{document}
\begin{titlepage}
\null

\begin{flushright}
KOBE-TH-96-04\\
hep-ph/9701302\\
December 1996
\end{flushright}

\vspace{7mm}
\begin{center}
  {\Large\bf Unitarity of Neutral Kaon System \par}
  \vspace{1.5cm}
  \baselineskip=7mm

 {\large Tsukasa Kawanishi\footnote{E-mail address: 
kawa@hetsun1.phys.kobe-u.ac.jp}\par}
\vspace{5mm}
 {\sl Graduate School of Science and Technology, Kobe University\\
     Rokkodai, Nada, Kobe 657, Japan \par}
  
\vspace{3cm}

{\large\bf Abstract}
\end{center}
\par

In neutral kaon system, we always use non-hermitian 
Hamiltonian for convenience of treating decay process, 
unitarity seems to be lost. If we take decay channels 
($\pi\pi$, $\pi\pi\pi$, $\pi\ell\nu$ $\ldots$ etc.) into 
account, however, Hamiltonian of the whole system must be 
hermitian. We attempt to derive an effective Hamiltonian 
with respect to only $K^0$, $\bar{K}^0$ states, starting from the
hermitian Hamiltonian. For brevity, we take only 
a $\pi\pi$ state into account as the decay channel in this paper.

We can not avoid an oscillation between $K^0$, $\bar{K}^0$
and $\pi\pi$ states if we start from a hermitian
Hamiltonian whose states all have discrete energy levels. 
We therefore treat the $\pi\pi$ state more appropriately
to have a continuous energy spectrum to achieve the decay 
of $K^0$, $\bar{K}^0$ into $\pi\pi$.

As the consequence, we find a different time evolution from what 
we expect in the conventional method immediately after the decay 
starts, though it recovers Fermi's golden rule for long enough 
time scale.

\end{titlepage}
\setcounter{footnote}{0}
\baselineskip=7mm

\sect{Introduction}

The neutral kaon system has long served as a probe of fundamental 
physics. Time evolution equation of neutral kaon system is 
written by non-hermitian Hamiltonian customarily, 
unitarity of the system seems to be lost. This is because
we don't take decay channels ($\pi\pi$, $\pi\pi\pi$, 
$\pi\ell\nu$ $\ldots$ etc.) of K mesons into account. 
Hamiltonian of the neutral kaon system must be hermitian with 
the decay channels being included into the base of 
this Hamiltonian. We attempt toderive the decay behavior of 
neutral kaon system effectively, starting from a hermitian 
Hamiltonian \cite{sanda}.

Recently, there are many discussions about CPT and quantum 
mechanics violation\cite{cpt}. If these violations exist, 
they must be extremely small. In order to treat 
these extremely small quantities, we need to describe 
the time evolution of neutral kaon system as exactly as possible. 
Therefore, it seems to be important to clarify whether 
the conventional treatment by use of non-hermitian Hamiltonian 
gives exact result, relying on the exact treatment. 
In future, such clarification may affect analysis of CPT 
and quantum mechanics violation.

In Sec.2, for convenience, we derive time evolution equation
of neutral kaon system with $K^0$, $\bar{K}^0$ and $\pi\pi$
bases. Here we treat $\pi\pi$ state as a state with 
continuous energy spectrum. Then we can avoid an oscillation 
between $\pi\pi$ and K meson, and explain 
a decay phenomenon well.

In Sec.3, assuming CP invariance, we solve the time evolution 
equation concerning $K_1$ and $\pi\pi$ states perturbatively, 
and consider the decay process of $K_1$.

In Sec.4, we examine the validity of making use of perturbation 
to solve time evolution equation. We introduce a simplified model
where we can solve the equation non-perturbatively, 
and compare the result with the one obtained by 
our perturbative method.

Sec.5 is devoted to summary and discussion.


\sect{Derivation of Time Evolution Equation }

  For brevity, we consider a system with $K^0$, $\bar{K}^0$
and $\pi\pi$ states. In a frame where the $K^0$, $\bar{K}^0$ are 
at rest, their 4-momenta are fixed, while $\pi\pi$ state has 
an ambiguity of relative momentum $\vec{k}$, 
$\pi(\vec{k})\pi( - \vec{k})$. If we start from $\pi\pi$ state as 
a discrete state, we can not avoid an oscillation 
$\pi\pi \leftrightarrow K$ . Thus $K^0$, $\bar{K}^0$ are
treated as discrete states, while $\pi\pi$ state is a continuous 
state with a continuous parameter $\vec{k}$.

At arbitrary time $ t $, a state $|\psi(t)>_I$ is given as
\begin{equation}
|\psi(t)>_I = C_{K^0} (t) |K^0> + C_{\bar{K}^0} (t) |\bar{K}^0>
+  \int {d^3}k C_{\vec{k}}(t) |\pi(\vec{k})\pi(-\vec{k})> ,
\end{equation}
where index $ I $ implies interaction representation.

We note that the momentum $\vec{P}$ should be preserved
in the $K^0$, $\bar{K}^0$ $\rightarrow$ $\pi\pi$ transition,
since the interaction Hamiltonian of the form
\begin{eqnarray}
& &{H^I}_{int} = \int {d^3}x {{\cal H}^I}_{int}, \\
& &{{\cal H}^I}_{int} = 
      \alpha K^0 \pi\pi + \bar{\alpha} \bar{K}^0 \pi\pi, \\
& &(\pi\pi : \pi^+ \pi^- , \pi^0 \pi^0  ; 
      CP invariance \rightarrow 
\alpha = - \bar{\alpha} ), \nonumber
\end{eqnarray}
implies that the total momentum should vanish due to the factor
$\delta (\vec{P_K} + \vec{P_\pi} + \vec{P_\pi})$. We thus have 
assigned $\vec{k} , - \vec{k}$ for the momentum of 
two $\pi$'s $(\vec{P_K} = \vec{0})$.

The normalization of continuous state is fixed as
\begin{equation}
<\pi (\vec{k}) \pi ( - \vec{k}) | \pi (\vec{k'}) \pi (- \vec{k'})>
= \delta^3 (\vec{k} - \vec{k'}).
\end{equation}
Then, $<\psi(t) | \psi(t)>_I = 1$ and Eq.(2.1), (2.4) gives
\begin{equation}
|C_{K^0} (t) |^2 + |C_{\bar{K^0}} (t) |^2 
      + \int {d^3}k |C_{\vec{k}} (t) |^2  = 1 
\end{equation}
Schr\"odinger equation is written as
\begin{eqnarray}
& &i\frac{\partial}{\partial t} |\psi(t)>_S = H|\psi(t)>_S \\
& &H = H_0 + H_{int}. \nonumber
\end{eqnarray}
Here, $ H_0 $ is free Hamiltonian and $ H_{int} $ is interaction
Hamiltonian due to weak interaction. Index $S$ implies 
Schr\"odinger representation.

It is related to the interaction picture by
\begin{eqnarray}
|\psi(t)>_I = e^{iH_0t} |\psi(t)>_S \\
i\frac{\partial}{\partial t} |\psi(t)>_I = {H^I}_{int} |\psi(t)>_I,
\end{eqnarray}
with the interaction Hamiltonian being defined by
\begin{equation}
{H^I}_{int} = e^{iH_0t} H_{int} e^{-iH_0t} .
\end{equation}
A "matrix" form of time evolution equation is possible by
inserting a complete set
\begin{equation}
1 = |K^0><K^0| + |\bar{K^0}><\bar{K^0}| + 
\int{d^3}k |\pi(\vec{k})\pi(-\vec{k})><\pi(\vec{k})\pi(-\vec{k})| ,
\end{equation}
to get
\begin{eqnarray}
\lefteqn{i\frac{\partial}{\partial t}\pmatrix{<K^0|\psi(t)>_I \cr 
<\bar{K^0}|\psi(t)>_I \cr 
  <\pi\pi|\psi(t)>_I \cr}} \nonumber \\ 
&=& \pmatrix{<K^0|{H^I}_{int}|K^0>& <K^0|{H^I}_{int}|\bar{K^0}>&
  <K^0|{H^I}_{int}|\pi\pi> \cr
   <\bar{K^0}|{H^I}_{int}|K^0>& <\bar{K^0}|{H^I}_{int}|\bar{K^0}>&
    <\bar{K^0}|{H^I}_{int}|\pi\pi> \cr
     <\pi\pi|{H^I}_{int}|K^0>&  <\pi\pi|{H^I}_{int}|\bar{K^0}>& 
      <\pi\pi|{H^I}_{int}|\pi\pi> \cr} \nonumber \\
& & \hspace{7cm} \times \pmatrix
  {<K^0|\psi(t)>_I \cr <\bar{K^0}|\psi(t)>_I \cr 
        <\pi\pi|\psi(t)>_I \cr}.
\end{eqnarray}
Eq.(2.1) and above gives
\begin{equation}
i\frac{\partial}{\partial t}
 \pmatrix{C_{K^0} (t) \cr C_{\bar{K^0}} (t) \cr 
     C_{\vec{k}} (t) \cr}  =
 \pmatrix{ {H^I}_{K^0K^0} &  {H^I}_{K^0{\bar{K^0}}} &
 {H^I}_{{K^0}{\vec{k'}}} \cr
 {H^I}_{\bar{K^0}K^0} & {H^I}_{\bar{K^0}\bar{K^0}} &
  {H^I}_{\bar{K^0}\vec{k'}} \cr
 {H^I}_{\vec{k}K^0} &  {H^I}_{\vec{k}\bar{K^0}} & 
  {H^I}_{\vec{k}\vec{k'}} \cr }
  \pmatrix{C_{K^0}(t) \cr C_{\bar{K^0}} (t) 
        \cr C_{\vec{k'}} (t) \cr}
\end{equation}
where multiplication of matrices should be understood as
\begin{equation}
 {H^I}_{K^0\vec{k}}  C_{\vec{k}} (t) \Rightarrow 
 \int {d^3}k  {H^I}_{K^0\vec{k}}  C_{\vec{k}}(t) .
\end{equation}
In Eq(2.12), the Hamiltonian in the interaction representation is
related to the time independent Hamiltonian in the Schr\"odinger
representation as
\begin{eqnarray}
& & \pmatrix{e^{iE_{K^0}t}& 0& 0 \cr
 0& e^{iE_{\bar{K^0}t}}& 0 \cr
  0& 0& e^{iE_{\vec{k}}t} \cr}
   \pmatrix{ H_{{K^0}{K^0}}&  H_{{K^0}{\bar{K^0}}}&
    H_{{K^0}{\vec{k'}}} \cr
     H_{\bar{K^0}{K^0}}& H_{\bar{K^0}\bar{K^0}}&
      H_{\bar{K^0}\vec{k'}} \cr
       H_{\vec{k}{K^0}}&  H_{\vec{k}\bar{K^0}}& 
        H_{\vec{k}\vec{k'}} \cr} \nonumber \\
& & \hspace{4cm} \times \pmatrix{ e^{-iE_{K^0}t} & 0 & 0 \cr
  0 & e^{-iE_{\bar{K^0}t}} & 0 \cr
   0 & 0 & e^{-iE_{\vec{k'}}t} \cr} 
  \nonumber \\
&=&\pmatrix{ H_{{K^0}{K^0}}&  H_{{K^0}{\bar{K^0}}}&
 e^{i \Delta E_{{K^0} \vec{k'}} t} H_{{K^0}{\vec{k'}}} \cr
 H_{\bar{K^0}{K^0}}& H_{\bar{K^0}\bar{K^0}}&
  e^{i \Delta E_{\bar{K^0} \vec{k'}} t} H_{\bar{K^0}\vec{k'}} \cr
 e^{i \Delta E_{\vec{k} {K^0}} t} H_{\vec{k} {K^0}}&  
 e^{i \Delta E_{\vec{k} \bar{K^0}} t} H_{\vec{k} \bar{K^0}}& 
 e^{i\Delta E_{\vec{k} \vec{k'}}t} H_{\vec{k} \vec{k'}} \cr}
\end{eqnarray}
where,
\begin{equation}
\Delta E_{\vec{k} \vec{k'}} = E_{\vec{k}} - E_{\vec{k'}},
\end{equation}
and so on.
CPT invariance implies
\begin{equation}
E_{K^0} = E_{\bar{K^0}} = M_{K}.
\end{equation}
%
%


\sect{Solving Time Evolution Equation Perturbatively}

In order to see how the unitarity is effectively lost 
we simplify the situation by assuming CP invariance, 
and consider only ($K_1, \pi\pi$) subsystem where $K_1$ 
denotes CP even eigenstate, i.e $K_2$ decouples from the system. 
We thus consider $K_1 \rightarrow \pi\pi $
decay and solve time evolution equation by perturbative method. 
For convenience, we treat $\pi$ as a massless particle.
Refering to Eq.(2.14), we obtain
\begin{eqnarray}
& &\pmatrix{C_{K_1}(t) \cr C_{\vec{k}} (t) \cr} = 
      {\bf T} e^{i \int_{0}^t \hat{H} (t)
dt} \pmatrix{C_{K_1} (0) \cr C_{\vec{k'}} (0) \cr} \nonumber \\
&=& (1 + i \int_{0}^t \hat{H} (t') dt' - 
      \int_{0}^t \hat{H} (t') dt' 
\int_{0}^{t'} \hat{H} (t") dt" + \cdots) \pmatrix
{C_{K_1} (0) \cr C_{\vec{k'}} (0) \cr} , \\
& &\hat{H}(t) = \pmatrix{0 & e^{i(E_{K_1} - E_{\vec{k'}} )t} 
      H_{{K_1}{\vec{k'}}} \cr
e^{-i(E_{K_1} - E_{\vec{k}})t} H_{\vec{k} K_1} & 0 \cr} ,
\end{eqnarray}
where $ \bf{T} $ stands for time ordered product and we have 
taken an interaction representation where
\begin{equation}
< K_1 | H_{int} | K_1> = 0 , \hspace{1cm} 
      <\pi\pi | H_{int} | \pi\pi > = 0.
\end{equation}
Inserting Eq.(3.2) into Eq.(3.1) and expanding in powers of 
$\hat{H} (t)$ to second order, we obtain
\begin{equation}
\pmatrix{C_{K_1}(t) \cr C_{\vec{k}} (t) \cr} = 
\pmatrix{ H_{11} & H_{12} \cr H_{21} & H_{22} \cr} 
\pmatrix{C_{K_1} (0) \cr C_{\vec{k'}} (0) \cr}, 
\end{equation}
\begin{equation}
 H_{11} = 1 + \int {d^3} k \frac{e^{i \Delta E_{{K_1} \vec{k}} t} 
 - 1 -i \Delta E_{{K_1} \vec{k}} t}{( \Delta E_{{K_1} \vec{k}} )^2}
    |H_{K_1 \vec{k}}|^2 
\end{equation}
\begin{equation}
 H_{12} = \frac{e^{i \Delta E_{{K_1} \vec{k'}} t} -1}
       {\Delta E_{{K_1} \vec{k'}}} H_{K_1 \vec{k'}}
\end{equation}
\begin{equation}
 H_{21} = - \frac{e^{-i \Delta E_{{K_1} \vec{k}} t} -1}
       {\Delta E_{{K_1} \vec{k}}} H_{\vec{k} K_1}
\end{equation}
\begin{equation}
H_{22} = 1 + \frac{1}{\Delta E_{{K_1} \vec{k'}}} 
( \frac{e^{i \Delta E_{\vec{k} \vec{k'}} t} -1}
      {\Delta E_{\vec{k} \vec{k'}}} 
+ \frac{e^{-i \Delta E_{{K_1} \vec{k}} t} -1}
      {\Delta E_{{K_1} \vec{k}}})
H_{\vec{k} K_1} H_{K_1 \vec{k'}} 
\end{equation}
and take initial conditions into consideration,
\begin{equation}
C_{K_1} (0) = 1 , \hspace{1cm} C_{\vec{k}} (0) = 0 .
\end{equation}
Then
\begin{equation}
|C_{K_1}(t)|^2 = \{1 + 2 \int d^3 k 
\frac{{\rm{cos}} (E_{K_1} - E_{\vec{k}} )t - 1}
      {(E_{K_1} - E_{\vec{k}} )^2}
 |H_{K_1 \vec{k}} |^2
+ O(H^4) \} |C_{K_1} (0) |^2 .
\end{equation}
Time derivative of Eq.(3.10) is
\begin{equation}
\frac{\partial}{\partial t} |C_{K_1}(t)|^2 = \{ -2 \int d^3 k 
\frac{{\rm{sin}} (E_{K_1} - E_{\vec{k}} )t}{E_{K_1} - E_{\vec{k}}} 
|H_{K_1 \vec{k}} |^2 + O(H^4) \} |C_{K_1} (t) |^2 .
\end{equation}
For the assumed interaction in Eq.(2.2) and Eq.(2.3), 
$|H_{K_1 \vec{k}} |^2 \propto 
      \frac{1}{\sqrt{E_{K_1}{E_{\vec{k}}}^2}}$.
Thus,
\begin{eqnarray}
\frac{\partial}{\partial t} |C_{K_1}(t)|^2 
&\propto& - \int_{M_{K_1}}^\infty dk \frac{{\rm{sin}} kt}{k} 
|C_{K_1} (t) |^2 \nonumber \\
&=& - \{ \pi - {\rm{si}} (M_{K_1} t)  \}
|C_{K_1} (t)|^2,
\end{eqnarray}
where si is sine integral function and $M_{K_1}$ denotes 
the $K_1$ mass. With $ t \rightarrow \infty $ , 
Eq.(3.12) reduces to what we expect from Fermi's golden rule,
\begin{equation}
\frac{\partial}{\partial t} |C_{K_1} (t)|^2
\propto - {\rm const} |C_{K_1}(t)|^2 .
\end{equation}
This equation indicates a decay process, not an  oscillation.
We should, however, note that when time $t$ is relatively small, 
our result shows a clear difference from the conventional result 
expected from the golden rule.


\sect{Comparison with a Non-perturbative Method}

In this section, we would like show that our procedure 
relying on a perturbative method actually reproduces 
the exact result solved nonperturbatively for some simplified case.
This indicates the validity of our method.

\subsection{A Simplified Model Solved by 
             Non-perturbative Method \cite{kawai}}

In this subsection, we treat the following system in 1+1 dimension.
\begin{eqnarray}
H | K_1 > = M_{K_1} | K_1 > + v \int dk | k >, \\
H | k > = v^* | K_1 > + k | k >,
\end{eqnarray}
where $k$ is assumed to take $-\infty$ to $\infty$, and $v$ 
is assumed to be constant.
We write eigenvalue and eigenvector of $H$ as $H | \omega > =
\omega | \omega >$ and difine $ | \omega >$ as following
\begin{equation}
| \omega > = N_{\omega} \{  | K_1 > 
      + \int dk f_{\omega} (k) | k > \}.
\end{equation}
Here, $N_{\omega}$ is a normalization factor. 
From the above we obtain
\begin{equation}
\omega = M_{K_1} + v^*\int dk f_{\omega} (k), \hspace{1cm} 
\omega f_{\omega} (k) = v + k f_{\omega} (k),
\end{equation}
which reads as
\begin{equation}
f_{\omega} (k) = \frac{{\bf P}}{\omega - k} 
      + \frac{\omega - M_{K_1} }{v^*} \delta (\omega -k).
\end{equation}
Here, {\bf P} denotes a principal value\cite{ww}. 
The normalization condition for continuous states,
\begin{equation}
< \omega | \omega ' > = \delta ( \omega - \omega ') , \hspace{1cm}
< k | k' > = \delta (k-k'),
\end{equation}
fixes $N_\omega$ as,
\begin{equation}
| N_{\omega} |^2 = \frac{|v|^2 }{ {\pi}^2 |v|^4 
      + (\omega - M_{K_1})^2}.
\end{equation}
At $t=0$, the state is assumed to be pure $K_1$ state, 
\begin{equation}
| \psi (0)> = |K_1 >= \int d \omega | \omega >< \omega | K_1 >= 
\int d \omega {N_{\omega}}^* | \omega>.
\end{equation}
Time evolution of $| \psi (t) >$ is uniquely fixed as
\begin{equation}
| \psi (t) >= \int d \omega e^{-i \omega t} 
      {N_{\omega}}^* | \omega >.
\end{equation}
The "survival probability" of $K_1$ is then given as
\begin{equation}
|<K_1 | \psi(t)>|^2 =
|\int d \omega \frac{|v|^2 e^{-i \omega t}}
      { {\pi}^2 |v|^4 + (\omega - M_{K_1})^2}|^2 \nonumber \\
= e^{-2 \pi |v|^2 t}.
\end{equation}
This expression means that a particle $K_1$ decays with  
$\Gamma = 2 \pi |v|^2$, and corresponds to the Fermi's golden rule. 
Namely, in this simplified model there is no deviation from 
the conventional result.

\subsection{Comparison with our Perturbative Model}

How about the result obtained by our perturbative method?. 
From Eq.(3.10), for 1+1 dimensional case we obtain 
($|H_{K_1 \vec{k}} |^2 = |v|^2$)
\begin{equation}
| C_{K_1} (t)|^2 = \{ 1+2 \int dk \frac{\cos (M_{K_1} - k)t -1}
      {(M_{K_1} - k)^2} |v|^2 + O( H^4 ) \} |C_{K_1} (0) |^2.
\end{equation}
Eq.(4.11) becomes
\begin{equation}
|C_{K_1} (t) |^2 = \{1-2 |v|^2 \pi t + O( |v|^4 ) \} 
      |C_{K_1} (0) |^2 .
\end{equation}
and time differential is 
\begin{equation}
\frac{ \partial }{ \partial t } 
      |C_{K_1} (t) |^2 \simeq -2 \pi |v|^2 |C_{K_1} (t)|^2 .
\end{equation}
This answer reproduces the exact result Eq.(4.10), and 
we realize that once we go through differential equation 
our method actually reproduces the nonperturbative 
result correctly.


\sect{Summary and Discussion}

In this paper, we have found a clear deviation from the result 
obtained by conventional Fermi's golden rule for relatively 
small $t$. But as is seen from Eq.(3.12) we can see this 
difference only for the time duration of order
$t \simeq \frac{1}{M_{K_1}}$. Therefore, it seems to be rather 
hard to observe such difference.
When, however, CPT symmetry is tested, such small deviation 
might affect the physical quantities, since CPT violation, 
if any, should be very tiny.


\vspace{1cm}
{\bf Acknowledgement}

I wish to record my special thanks to A.I.Sanda(Nagoya-U) 
for stimulating discussing and M.Kobayashi(KEK), C.S.Lim (Kobe-U) 
and T.Morozumi (Hiroshima-U) for their suggestion, and H.Kawai 
(KEK) and H.Sonoda (UCLA) for explaining the prescription
in subsection 4.1 .



\end{document}